# Impact of Link Failures on the Performance of MapReduce in Data Center Networks

Sanaa Hamid Mohamed, Taisir E. H. El-Gorashi, and Jaafar M. H. Elmirghani
*School of Electronic & Electrical Engineering, University of Leeds, LS2 9JT, United Kingdom*

**ABSTRACT**
In this paper, we utilize Mixed Integer Linear Programming (MILP) models to determine the impact of link failures on the performance of shuffling operations in MapReduce when different data center network (DCN) topologies are used. For a set of non-fatal single and multi-links failures, the results indicate that different DCNs experience different completion time degradations ranging between 5% and 40%. The best performance under links failures is achieved by a server-centric PON-based DCN.

## 1. INTRODUCTION

Providing high reliability and continuity is a key requirement for cloud computing services as any disruption or disconnection in the infrastructure typically leads to revenue losses and customer departures. Failures in cloud infrastructure can occur in the transporting networks, data centers, or in the applications [1]. Overcoming the severe impacts of these failures requires the adoption of resilient designs and restoration plans. The resilience and energy efficiency of different IP over WDM networks topologies were examined in [2] under fiber cuts or a core node failure. The work in [3] and [4] considered static and dynamic adaptation to traffic surges resulting from re-routing demands after links failures. In [5], an energy efficient 1+1 protection scheme in IP over WDM networks was proposed based on Network Coding (NC) while considering the impact of the topology and the traffic. The authors in [6] addressed the resilience of geo-distributed transport networks that link cloud data centers by jointly considering the content placement, and anycast routing under failures. At the application level, solutions such as maintaining several replicas of data and software components are considered. For example, MapReduce, which is a widely used framework for big data parallel computations, has a data replication mechanism with default value of 3, and considers speculative execution for straggler tasks [7]. Several studies have considered the resilience of Data Center Networks (DCN) (e.g. [8]-[11]) by addressing the sources of links, switches, and servers' software and hardware failures.

In this paper, we examine single and multi-link non-fatal failures in several DCNs and their effect on the performance of MapReduce shuffling operations. We compare the completion time results in [12] and [13] under no link failures to the results under several link failures for electronic, hybrid, and optical DCNs, in addition to a server-centric Passive Optical Networks (PON)-based DCN design proposed as part of the work in [14]-[19]. The rest of this paper is organized as follows. Section 2 describes the proposed optimization models and shows their results, while Section 3 provides the conclusions and future work.

## 2. OPTIMIZING SHUFFLING OPERATIONS IN DATA CENTERS UNDER LINKS FAILURES

### 2.1 Methodology

We utilize Mixed Integer Linear Programming (MILP) models to examine the effects of single and multi-link failures in different DCNs on the optimum routing of intermediate data while balancing the completion time and the energy efficiency. Seven DCNs [12] namely Spine-and-Leaf, Fat-Tree, BCube, DCell, c-Through, Helios, in addition to the server-centric PON-based DCN design in [16] with 4 racks and 1 server per group, are considered. Figure 1 illustrates the assignment of map and reduce workers in 10 and 6 servers, respectively in each DCN model. It also shows the link disconnection cases considered. Indy GraySort benchmark [20] is selected as it is a representative workload for examining the congestion in DCNs due to routing intermediate data that is equal in size to the input data from the 10 map workers to the 6 reduce workers. The power consumption values of the selected electronic and optical equipment are as in [12], and [15].

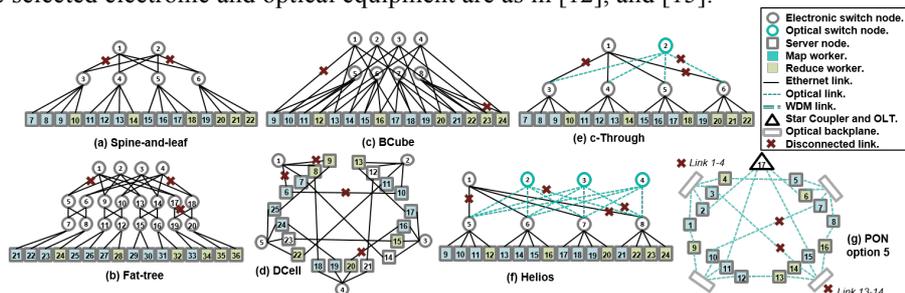

*Figure 1: DCN models considered for MapReduce shuffling optimization under different links failures.*

## 2.2 Results

Figure 2 provides the shuffling completion time results for a maximum server rate of 1000 Mbytes/s and input data ranging from 1 to 20 GBytes in different DCNs. The subfigures, show the results in [12], and [13] for the cases where all links are working, in addition to the completion time results under 2 or 3 cases of links failures in each DCN. For the Spine-and-Leaf architecture, the degradation due to links 1-3 disconnection is less than 2-6 because the latter is more utilized due to serving the flows to 3 reduce workers compared to 1 reduce worker by links 1-3. In the Fat-Tree architecture, we examined the impact of disconnecting the most utilized links (i.e. 17-20). Any additional links disconnections did not deteriorate the performance. In both DCNs, failure in any link between the Top of Rack (ToR) and a server is considered a single point-of-failure and requires a software-based resilience mechanism such as replication, where the map computations should be re-performed in another server containing a replica. The BCube and DCell architectures are server-centric DCNs with two links to each server which provides more routes compared to the Spine-and-Leaf architecture and the Fat-Tree architecture, and hence, are more resilient to link failures. Single and multi-link failures including the most utilized link in BCube caused an increase of 24% in the completion time. The DCell architecture exhibited the highest degradation in the completion time (i.e. 40%) with links 1-6 disconnection which are the most utilized due to serving two reduce workers from other racks, which indicates high sensitivity to workload placements. An average increase by 37% in the completion time is experienced in the c-Through architecture and by 5% in the Helios architecture for three link failures because more optical switches are utilized for the same number of ToR switches. Both however have a single-point-of-failure as in the Fat-Tree and Spine-and-Leaf architectures. The server-centric PON-based design exhibited an average degradation of 22% under single and up to 4 link failures. Figure 3 summarizes the overall completion time (i.e. completion time without failures in addition to the extra delay due to failures) for the cases considered in Figure 2.

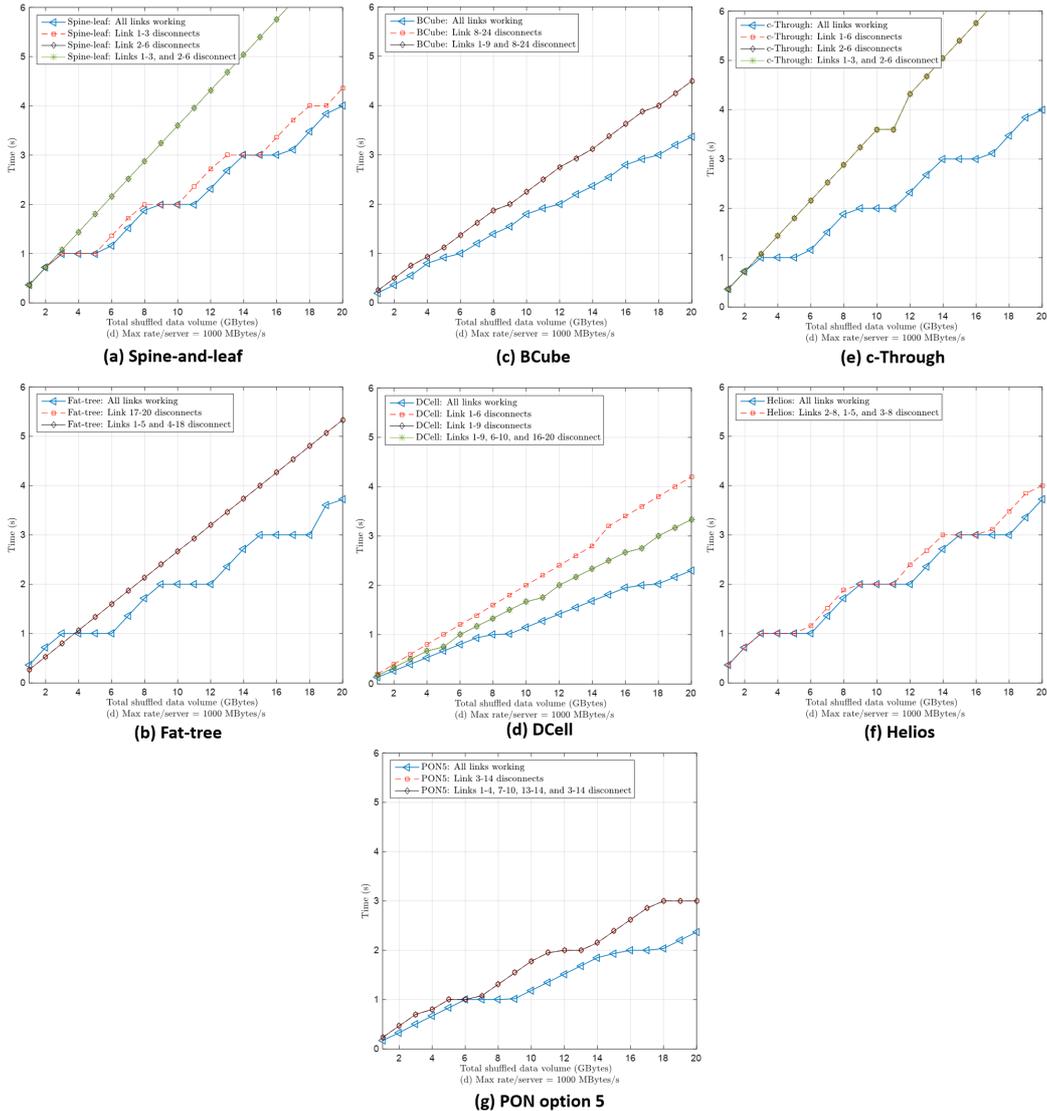

*Figure 2: Shuffling completion time in different DCNs under links failure for max. rate/server of 1000 MBytes/s.*

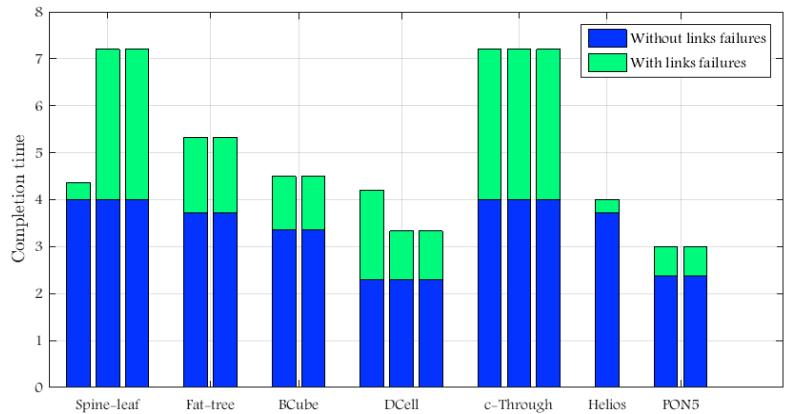

*Figure 3: Overall completion time for total shuffled data volume of 20 GBytes.*

## 3. CONCLUSIONS AND FUTURE WORK

In this paper, we examined the impact of single and multi-link failures in different DCN topologies under extensive server-to-server communications. The results for a set of non-fatal single and multi-link failures indicate that different DCNs experience different degradations based on the links redundancy, workloads placement, and links utilizations. The best overall performance was achieved by the server-centric PON-based DCN in [16]. Future work includes considering the switching delay, switches failures, and the data replication factor.

## ACKNOWLEDGEMENTS

The authors would like to acknowledge funding from the Engineering and Physical Sciences Research Council (EPSRC), through INTERNET (EP/H040536/1) and STAR (EP/K016873/1) projects. The first author would like to acknowledge EPSRC for funding her PhD programme of study. All data are provided in full in the results section of this paper.